\documentclass[conference]{IEEEtran}
\IEEEoverridecommandlockouts

\usepackage{cite}
\usepackage{amsmath,amssymb,amsfonts}
\usepackage{algorithmic}
\usepackage{graphicx}
\usepackage{url}
\usepackage{textcomp}
\usepackage[table]{xcolor}
\usepackage{booktabs}
\usepackage{amsmath,amsfonts}
\usepackage{multirow}
\usepackage{arydshln}
\usepackage{rotating}
\usepackage{fdsymbol}
\usepackage[T1]{fontenc}
\usepackage[framemethod=TikZ]{mdframed}
\definecolor{revised}{RGB}{0, 0, 255}
\newcommand{\myccone}{\cellcolor[HTML]{f2f2f2}}
\newcommand{\mycctwo}{\cellcolor[HTML]{d9d9d9}}

\begin{document}

\title{~\includegraphics[height=20pt]{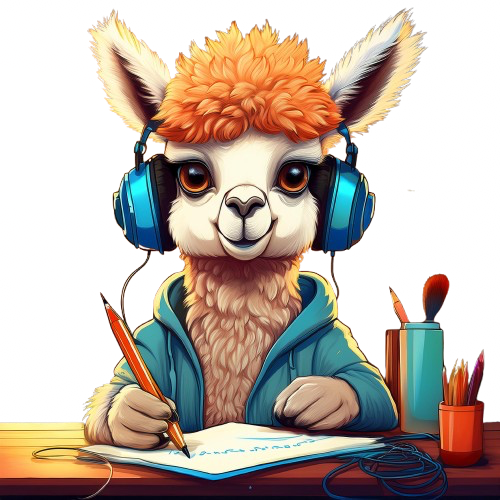} ReCLAP: Improving Zero Shot Audio Classification by Describing Sounds}

\author{\IEEEauthorblockN{Sreyan Ghosh$^{\spadesuit}$ \quad Sonal Kumar$^{\spadesuit\vardiamondsuit}$ \quad Chandra Kiran Reddy Evuru $^{\vardiamondsuit}$ \quad Oriol Nieto$^{\vardiamondsuit}$\\ Ramani Duraiswami$^{\vardiamondsuit}$ \quad Dinesh Manocha$^{\vardiamondsuit}$}
\IEEEauthorblockA{$^{\spadesuit}$University of Maryland, College Park, MD, USA \quad $^{\vardiamondsuit}$Adobe Research, San Francisco, CA, USA \\
\{sreyang, sonalkum, ckevuru, ramanid, dmanocha\}@umd.edu}
}


\maketitle

\begin{abstract}
Open-vocabulary audio-language models, like CLAP~\cite{elizalde2023clap}, offer a promising approach for zero-shot audio classification (ZSAC) by enabling classification with any arbitrary set of categories specified with natural language prompts. In this paper, we propose a simple but effective method to improve ZSAC with CLAP. Specifically, we shift from the conventional method of using prompts with abstract category labels (e.g.,~\textit{Sound of an organ}) to prompts that describe sounds using their inherent descriptive features in a diverse context (e.g.,~\textit{The organ's deep and resonant tones filled the cathedral.}). To achieve this, we first propose \textbf{ReCLAP}, a CLAP model trained with \underline{re}written audio captions for improved understanding of sounds in the wild. These rewritten captions describe each sound event in the original caption using their unique discriminative characteristics. ReCLAP outperforms all baselines on both multi-modal audio-text retrieval and ZSAC. Next, to improve zero-shot audio classification with ReCLAP, we propose \textit{prompt augmentation}. In contrast to the traditional method of employing hand-written template prompts, we generate custom prompts for each unique label in the dataset. These custom prompts first describe the sound event in the label and then employ them in diverse scenes. Our proposed method improves ReCLAP's performance on ZSAC by 1\%-18\% and outperforms all baselines by 1\% - 55\%\footnote{Code and Checkpoints: \url{https://github.com/Sreyan88/ReCLAP}}.
\end{abstract}

\section{Introduction}

Audio classification, the foundational task of assigning a category label to an audio sample, remains one of the most important tasks in audio processing  and has numerous real-world applications. Zero-shot audio classification (ZSAC) presents a promising approach that provides greater flexibility at the inference stage than supervised methods. Unlike supervised methods that map input audio to a fixed set of categories, models classify by computing a similarity score between an input audio example and a caption. To perform inference, one can generate a caption or ``prompt'' associated with each desired category and match each audio sample to the best prompt. This means that categories can be selected ad hoc and adjusted without additional training.

Open-vocabulary audio-language models like Contrastive Language-Audio Pre-training (CLAP) \cite{elizalde2023clap} have outperformed most other models on ZSAC. CLAP is trained on contrastive objectives between audio-caption pairs, where each audio sample corresponds to non-speech sounds and non-verbal speech, and the captions describe the acoustic events and the scene, not the spoken content. Beyond ZSAC, CLAP has also achieved superior performance on cross-modal audio-text retrieval~\cite{ghosh2024compa} and has been used as a backbone audio encoder for a variety of audio-language tasks, including generalist audio agents~\cite{deshmukh2023pengi}, open-domain chat assistants~\cite{kong2024audio}, audio captioning~\cite{ghosh2023recap} and text-to-audio generation~\cite{liu2023audioldm}. However, ZSAC with CLAP currently remains subpar compared to standard supervised methods~\cite{silva2023collat}. We attribute this to 3 main reasons:
\begin{enumerate}
    \item \textbf{Limited access to large-scale audio-caption datasets}: Unlike CLIP~\cite{radford2021learning}, CLAP has not been trained on large-scale, open-source audio-caption datasets~\cite{ghosh2024compa}. This constrains its ability to fully understand and perceive the diverse range of audio and language interactions~\cite{ghosh2024compa}.
    \item \textbf{Lack of generalization beyond training category labels}: CLAP struggles to generalize beyond the specific category labels used in its training prompts. For instance, research by Tiago \textit{et al.}~\cite{tavares2024class} indicates that a model’s ZSAC accuracy is closely related to the clusters in its audio embedding space. Consequently, if CLAP was trained on a dataset where the prompt is \textit{“Sound of a toothbrush”} from AudioSet~\cite{gemmeke2017audio}, it might not accurately generalize to a similar label like “brushing teeth” in the ESC50 dataset~\cite{piczak2015dataset}, even though the sounds are similar (\textit{both sound like a soft scrubbing or swishing noise, often with a light, scratchy texture}).
    \item \textbf{Limitations of hand-written prompts for ZSAC}: The current ZSAC setup relies on hand-written prompts that correspond directly to dataset category labels. These prompts fail to provide additional context beyond the label itself. For example, CLAP may struggle to classify a label like ``Residential Area'' in the CochlScene dataset~\cite{jeong2022cochlscene} if it has not encountered that label during training. The label alone offers very little information about what sounds characterize a residential area, leading to potential misclassification.
\end{enumerate}

\vspace{0.5mm}



\begin{figure}[t]
    \centering
    \includegraphics[width=\columnwidth]{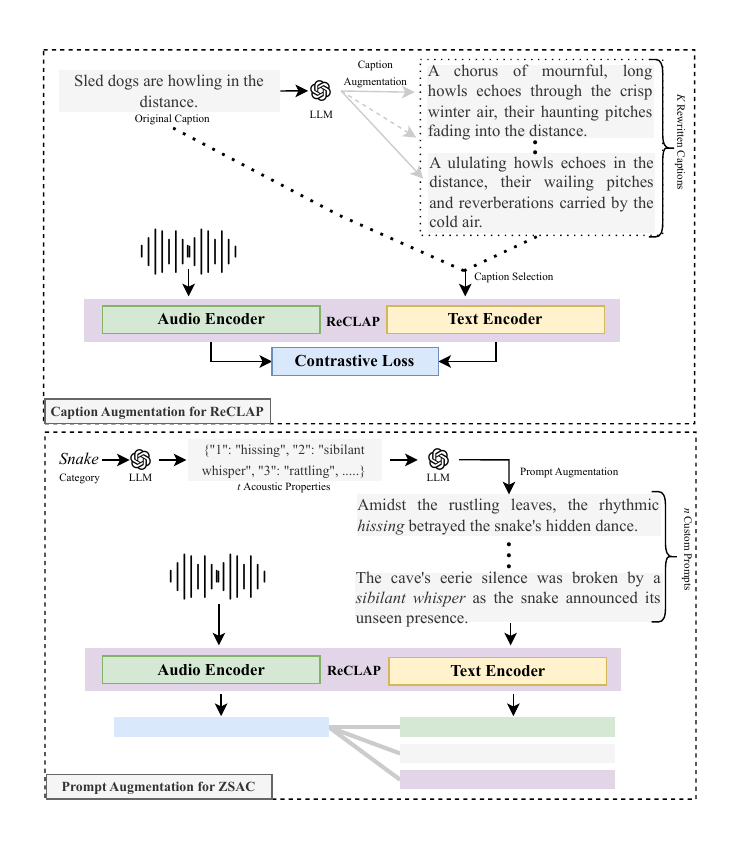}
    \caption{\small Illustration of our proposed method for improving Zero Shot Audio Classification (ZSAC) with language augmentation. \textbf{Top:} We enhance CLAP training through \textit{caption augmentation}, where each audio's caption is expanded and rewritten by prompting LLMs to provide detailed descriptions of the sound events. During training, we choose either the original caption or one of the rewritten captions. \textbf{Bottom:} We perform \textit{prompt augmentation} and generate custom prompts for each label category in the dataset. These prompts describe the sound in the category in diverse scenes.}
    \label{fig:enter-label}
    \vspace{-1.3em}
\end{figure}

{\noindent \textbf{Main Contributions.}} In this paper, we propose a simple, scalable, and effective approach to improve ZSAC with CLAP. Our contributions are twofold and are summarized as follows:
\begin{enumerate}
    \item We present ReCLAP, a CLAP model trained using \textit{caption augmentation}. Specifically, we prompt a Large Language Model (LLM) to generate multiple diverse rewrites of the caption associated with each audio. Each rewrite describes the sounds in a unique way. Additionally, they exhibit diversity in sentence structure and vocabulary while preserving the original key concepts and meanings (example in Fig.~\ref{fig:enter-label} and Section~\ref{subsec:training}). This simple data augmentation technique has several advantages, including \textbf{(1)} It enables the model to learn about the distinct acoustic features of sound events beyond what abstract labels alone can provide. This leads to more accurate clustering of sounds based on their actual acoustic properties rather than relying solely on predefined labels. \textbf{(2)} Text-based augmentation via LLM-generated captions provides an effective and scalable method for training-time data augmentation. Unlike traditional data augmentation techniques, which typically involve random audio perturbations, our method is more interpretable and avoids the complexities and limitations of generating synthetic audio. ReCLAP achieves state-of-the-art performance across various retrieval, and ZSAC benchmarks with standard setups.
    
    \item To further improve ZSAC performance with ReCLAP, we go beyond simple hand-written template prompts (e.g., “The sound of a \{category\}”) by generating multiple custom prompts for each category. This process involves two steps: \textbf{(1)} We prompt an LLM to describe the sound of each category label in $t$ distinct ways, focusing on its unique acoustic characteristics (e.g., Gasp: “a sharp intake of breath”). \textbf{(2)} We then create prompts that place the sound event in diverse scenes, incorporating the descriptions generated in the previous step (e.g., “A sharp intake of breath sliced through the silence as the verdict was announced”). Our proposed method improves the performance of ReCLAP across various ZSAC benchmarks by 1\%-55\%.
\end{enumerate}

\section{Related Work}
\label{sec:related}

Following the initial work on CLAP~\cite{elizalde2023clap}, several works have worked to improve its performance. For example, Wu \textit{et al.}~\cite{laionclap2023} scaled CLAP to 630k audio-caption pairs (including proprietary datasets) and showed a considerable boost in performance. Following this, Elizade \textit{et al.} scaled their data to 4.6M audio-caption pairs and included speech samples in their training. Ghosh \textit{et al.}~\cite{ghosh2024compa} employed 660k pairs using only public domain data to build CompA-CLAP. They also proposed novel techniques to improve the compositional reasoning abilities of CLAP. CLAP has also been employed as an audio or text backbone for a variety of foundational audio processing tasks including text-to-audio generation~\cite{ghosal2023tango,liu2023audioldm,huang2023make,agostinelli2023musiclm}, audio captioning~\cite{ghosh2023recap} and audio chat models~\cite{gong2024listen,kong2024audio,deshmukh2023pengi}. Despite its gaining popularity, research efforts to enhance CLAP's audio and language comprehension skills have been limited, with prior work focusing mainly on scaling.

\section{Methodology}
\label{sec:methodology}


\subsection{Caption Augmentation for ReCLAP}
\label{subsec:training}

CLAP is trained on a contrastive objective between audio-caption pairs to learn a shared representation between the audio and language modalities. Specifically, let $X_a$ and $X_t$ be the audio and its corresponding caption. Additionally, let $f_a(.)$ and $f_b(.)$ be the audio and text encoders respectively. We first obtain audio and text representations $\hat{X}_a$ and $\hat{X}_b$ as follows:
\begin{equation}
    \hat{X}_a=f_a\left(X_a\right) ; \hat{X}_t=f_t\left(X_t\right)
\end{equation}
where $\hat{X}_a \in \mathbb{R}^{N \times D}$ and $\hat{X}_b \in \mathbb{R}^{N \times D}$. $N$ here is the batch size and $D$ is the embedding dimension. Next, we measure similarity as follows:
\begin{equation}
C=\tau *\left(\hat{X}_a \cdot \hat{X}_t^{\top}\right)
\label{eqtn:similarity}
\end{equation}
where $C$ is any similarity function that measures distance using the dot product (cosine similarity in our case). Finally, the contrastive loss is calculated as:
\begin{equation}
    \mathcal{L}=0.5 *\left(\ell_{\text {text }}(C)+\ell_{\text {audio }}(C)\right)
\end{equation}
\begin{equation}
    \ell_k=\frac{1}{N} \sum_{i=0}^N \log \operatorname{diag}(\operatorname{softmax}(C))
\end{equation}
We train ReCLAP with a training objective similar to that in CLAP but with \textit{caption augmentation}. Specifically, we augment each training sample with $K$ additional text captions by rewriting the original caption associated with each audio sample in the dataset in $K$ diverse ways. During training, for each audio sample, ReCLAP chooses the original caption with a probability $p = 0.4$ or one of the rewritten versions (with a probability $1-p$) where each rewritten caption has an equal probability of selection. Thus, Eqn.~\ref{eqtn:similarity} can be re-written as:

\begin{equation}
C=\tau *\left(\hat{X}_a \cdot f_t(\operatorname{aug}_t(X_t))^{\top}\right)
\label{eqtn:similarity2}
\end{equation}

where $\operatorname{aug}_t(.)$ denotes the rewriting and choosing operation. The primary objective of the rewriting operation is to rewrite the caption so that each sound in the caption is described using its unique acoustic characteristics. An example is as follows:


\begin{mdframed}[linewidth=1pt, linecolor=black, leftmargin=1pt, rightmargin=1pt, innerleftmargin=10pt, innerrightmargin=10pt, innertopmargin=4pt, innerbottommargin=2pt, backgroundcolor=gray!20, roundcorner=5pt]
\textit{\textbf{(1) Original Caption:}} A traction engine is idling.

\noindent\textit{\textbf{(1) Rewritten Caption:}} A low, rumbling diesel engine hums steadily, its vibrations resonating through the air.

\noindent\textit{\textbf{(2) Original Caption:}} Cars are starting in pairs.

\noindent\textit{\textbf{(2) Rewritten Caption:}} Rapid, low-pitched revving of engines, followed by the synchronized, high-pitched roar of multiple cars starting in unison.

\end{mdframed}

We instruct an LLM to complete this task, which ensures that the rewritten captions or augmentations exhibit high levels of diversity in sentence structure and vocabulary while preserving the original key concepts and meanings. The instruction used to prompt the LLM is provided in our GitHub. We employ LLaMa-3.1-8B~\cite{touvron2023llama} with in-context examples written by humans. We randomly sample 5 in-context examples for every prompt from a collection of 50.

\subsection{Prompt Augmentation for ZSAC}
\label{subsec:custom}
\begin{table*}[t]
\centering
\caption{\small Performance comparison of ReCLAP with baselines on Text-to-Audio and Audio-to-Text retrieval on AudioCaps and Clotho. ReCLAP outperforms baselines by 0.4\%-38.9\%.}
\resizebox{2\columnwidth}{!}{
\begin{tabular}{lccc|ccc|ccc|ccc}
\toprule \toprule
\multicolumn{1}{c}{} &
  \multicolumn{6}{c|}{\textbf{AudioCaps}} &
  \multicolumn{6}{c}{\textbf{Clotho}} \\
\multicolumn{1}{c}{Model} &
  \multicolumn{3}{c}{Text-to-Audio} &
  \multicolumn{3}{c|}{Audio-to-Text} &
  \multicolumn{3}{c}{Text-to-Audio} &
  \multicolumn{3}{c}{Audio-to-Text} \\
  & R@1 & R@5 & R@10 & R@1 & R@5 & R@10 & R@1 & R@5 & R@10 & R@1 & R@5 & R@10 \\ 
  \midrule
MMT  & 36.1 & 72.0 & \underline{84.5} & 39.6 & 76.8 & 86.7 & 6.7 & 21.6 & 33.2 & 7.0 & 22.7 & 34.6\\
ML-ACT  & 33.9 & 69.7 & 82.6 & 39.4 & 72.0 & 83.9 & 14.4 & 36.6 & 49.9 & 16.2 & 37.6 & 50.2 \\
CLAP      & 34.6 & 70.2 & 82.0 & 41.9 & 41.9 & 84.6 & 16.7 & 41.1 & 54.1 & \underline{20.0} & 44.9 & 58.7 \\
CompA-CLAP & 36.1 & 72.6 & 81.6 & {45.2} & \underline{80.1} & 86.7 & 16.8 & 43.5 & {56.1} & {19.7} & \underline{45.2} & 55.6 \\ 
LAION-CLAP (repro.) & 34.5 & 70.7 & 80.2 & 42.5 & 77.9 & {87.4} & {15.8} & 39.7 & 52.9 & 19.1 & 44.1 & 54.9 \\ 
 \midrule
\myccone CLAP-2.3M &\myccone \underline{36.2} &\myccone \underline{72.7} &\myccone 82.8 &\myccone \underline{46.1} &\myccone {79.1} &\myccone {87.5} &\myccone \underline{17.1} &\myccone 43.9 &\myccone \underline{56.8} &\myccone {19.2} &\myccone {41.2} &\myccone {56.4} \\
\myccone ReCLAP-660k  &\myccone {35.9} &\myccone 72.3 &\myccone 82.5 &\myccone {45.2} &\myccone {79.6} &\myccone 87.9 &\myccone 16.8 &\myccone \underline{44.1} &\myccone 55.8 &\myccone {18.9} &\myccone {42.8} &\myccone \underline{57.3} \\
\myccone ReCLAP &\myccone \textbf{37.1} &\myccone \textbf{73.2} &\myccone \textbf{85.0} &\myccone \textbf{48.0} &\myccone \textbf{80.4} &\myccone \textbf{90.8} &\myccone \textbf{18.9} &\myccone \textbf{44.7} &\myccone \textbf{59.0} &\myccone \textbf{20.5} &\myccone \textbf{45.7} &\myccone \textbf{58.9} \\ \bottomrule
\end{tabular}}
\label{tab:exp-t2a-retrieval}
\end{table*}
After training ReCLAP with rewritten captions, we can now employ ReCLAP for ZSAC. However, the standard approach for ZSAC is to handwrite a prompt template and use it for every category in the classification dataset (e.g.,\textit{``The sound of a \{category\}''}), like ~\cite{laionclap2023,elizalde2023clap}. However, this method has a major drawback: The prompt merely specifies the category without providing details about the unique acoustic characteristics of the audio concept corresponding to the category. This limits CLAP's understanding of arbitrary categories in the wild, which are just abstract definitions of audio concepts. Therefore, incorporating descriptions of a category's acoustic features into the prompt provides CLAP with an intermediate level of understanding regarding the expected sound of that category. Our proposed method also complements ReCLAP, which possesses additional knowledge about the acoustic features of many audio concepts.

Thus, moving from one standard prompt for every category, we propose employing $N$ custom prompts for every category in the dataset. Since manually hand-writing such custom prompts is infeasible, we instruct an LLM for this task. We instruct an LLM in two stages, with two different instructions. In the first stage, we instruct an LLM to describe the sound of each category label in $t$ distinct ways,
focusing on its unique acoustic characteristics:

\begin{mdframed}[linewidth=1pt, linecolor=black, leftmargin=1pt, rightmargin=1pt, innerleftmargin=10pt, innerrightmargin=10pt, innertopmargin=4pt, innerbottommargin=2pt, backgroundcolor=gray!20, roundcorner=5pt]

\noindent\textit{\textbf{(1) Category:}} Bicycle bell (FSD50k)

\noindent\textit{\textbf{(1) Acoustic Properties:}} (i) metallic ring, (ii) high-pitched, tinkling chime $\cdots$

\noindent\textit{\textbf{(2) Category:}} mallet (NSynth)

\noindent\textit{\textbf{(2) Acoustic Properties:}} (i) dull thud, (ii) resonant knock (iii) deep thump, $\cdots$
\end{mdframed}

Next, using these descriptions, we instruct an LLM to generate $n$ different captions for each property, with the sound described in the category occurring in diverse scenes:

\begin{mdframed}[linewidth=1pt, linecolor=black, leftmargin=1pt, rightmargin=1pt, innerleftmargin=10pt, innerrightmargin=10pt, innertopmargin=4pt, innerbottommargin=2pt, backgroundcolor=gray!20, roundcorner=5pt]
\noindent\textit{\textbf{(1) Prompt Caption:}} A bicycle bell's clear, \underline{metallic ring} slices the silence as a rider announces their presence in the peaceful park.

\noindent\textit{\textbf{(2) Prompt Caption:}} The mallet's \underline{dull thud} reverberated through the silent courtroom as the judge announced the verdict.
\end{mdframed}

Finally, for every category, we randomly sample $N$ unique prompts from the pool of $n \times t$ total prompts. For ZSC, we mean pool $N$ text embeddings corresponding to the prompts for every label ($\mathbb{R}^{N \times d} \rightarrow \mathbb{R}^{d}$, where $d$ is the shape of embedding output by CLAP). Finally, we calculate the cosine similarity between each audio embedding and all text embedding for all the labels for ZSAC.




\section{Experimental Setup}
\label{sec:experimental_setup}

{\noindent \textbf{ReCLAP Training Datasets.}} We train ReCLAP from scratch on a collection of multiple datasets including Sound-VECaps~\cite{yuan2024improving} and CompA-660k~\cite{ghosh2024compa}. Detailed statistics about each dataset are provided on our GitHub. Our dataset has $\approx$2.3M audio-caption pairs.
\vspace{1mm}

{\noindent \textbf{Evaluation Datasets.}} For ZSAC, we adopt an evaluation setup similar to prior works~\cite{laionclap2023,elizalde2023clap,ghosh2024compa} and employ AudioSet~\cite{gemmeke2017audio}, ESC-50~\cite{piczak2015dataset}, FSD50k~\cite{fonseca2022fsd50k}, NSynth~\cite{engel2017neural}, TUT-Urban~\cite{mesaros2018multi}, UrbanSound8K~\cite{salamon2017deep} and VGGSound~\cite{Chen20}. We evaluate for accuracy on multi-class and mAP for multi-label.
\vspace{1mm}

{\noindent \textbf{Model Architecture and Hyper-parameters.}} We follow the same model architecture as  CompA-CLAP~\cite{ghosh2024compa} with a T5~\textsubscript{large} text encoder~\cite{2020t5} and HTSAT~\textsubscript{base} audio encoder~\cite{htsat-ke2022}. We train ReCLAP with a learning rate of 5e-4 and an effective batch size (BS) of 256. This BS is smaller than that in the literature, but we do so due to computational constraints. We employed $k$=4 for caption and $N$ = 2 for prompt augmentation. 
\vspace{1mm}

\begin{table*}[]
\caption{\small Performance comparison of ReCLAP with baselines on Zero-shot Audio classification benchmarks. ReCLAP outperforms baselines by 0.6\%-54.8\%.}
    \centering
    \resizebox{1.5\columnwidth}{!}{
\begin{tabular}{llccccccc}
\toprule \toprule
\multicolumn{2}{c}{} &  ESC-50 &  US8K &  VGGSound &  FSD50K &  TUT &  AudioSet &  NSynth \\ \midrule
\multirow{7}{*}{{w/o Prompt Aug.}} &  Wav2CLIP &  41.4 &  40.4 &  10.0 &  3.0 &  28.2 &  5.0 &  5.9 \\
 &  AudioClip &  69.4 &  65.3 &  9.9 &  6.6 &  29.5 &  3.7 &  6.8 \\
 &  CLAP &  82.6 &  73.2 &  16.4 &  14.0 &  29.6 &  5.1 &  9.9 \\
 &  LAION-CLAP (repro.) &  88.2 &  74.1 &  21.2 &  22.4 &  58.4 &  20.8 &  11.8 \\
 &  CoLLAT &  84.0 &  77.0 &  - &  19.0 &  29.0 &  9.0 &  - \\
 &  CompA-CLAP &  86.5 &  88.1 &  21.9 &  19.6 &  56.7 &  21.6 &  11.8 \\
 &  \myccone CLAP-2.3M &  \myccone 88.6 &  \myccone 90.3 &  \myccone 24.5 &  \myccone {30.6} &  \myccone 61.5 &  \myccone 21.9 &  \myccone 11.1 \\
\arrayrulecolor{black}\cdashline{2-9} 
&  \myccone ReCLAP &  \myccone 90.0 &  \myccone 94.3 &  \myccone 24.7 &  \myccone 27.2 &  \myccone 63.3 &  \myccone 23.5 &  \myccone 11.4 \\ \bottomrule
\multirow{6}{*}{{w/ Prompt Aug.}} &  CLAP &  83.4 &  74.5 &  16.4 &  14.9 &  33.7 &  6.2 &  10.2 \\
 &  LAION-CLAP (repro.) &  89.5 &  76.3 &  23.1 &  24.5 &  61.5 &  21.4 &  12.4 \\
 &  \myccone CLAP-2.3M &  \myccone 89.9 &  \myccone 91.2 &  \myccone 25.2 &  \myccone \underline{37.8} &  \myccone 63.7 &  \myccone 23.2 &  \myccone 13.1 \\
 &  \myccone ReCLAP-660k &  \myccone 89.5 &  \myccone 79.0 &  \myccone 25.9 &  \myccone 28.9 &  \myccone 60.3 &  \myccone 22.9 &  \myccone 13.6 \\
 &  \myccone ReCLAP w/ only desc. &  \myccone \underline{89.6} &  \myccone \underline{92.9} &  \myccone \underline{26.8} &  \myccone {37.1} &  \myccone \underline{65.9} &  \myccone \underline{25.4} &  \myccone \underline{14.1} \\
 \arrayrulecolor{black}\cdashline{2-9}
 &  \mycctwo ReCLAP &  \mycctwo \textbf{92.8} &  \mycctwo \textbf{95.2} &  \mycctwo \textbf{29.2} &  \mycctwo \textbf{40.2} &  \mycctwo \textbf{67.4} &  \mycctwo \textbf{26.1} &  \mycctwo \textbf{14.7} \\ \hline
\end{tabular}}
    \label{tab:zshot_res}
\end{table*}

{\noindent \textbf{Baselines.}} We use the following baselines for comparison: MMT \cite{oncescu2021audio}, ML-ACT~\cite{mei2022metric}, CLAP~\cite{elizalde2023clap}, CompA-CLAP~\cite{ghosh2024compa}, LAION-CLAP~\cite{laionclap2023} and LAION-CLAP~\textit{(ours)} (reproduced with BS=256 and excluding non-open-source datasets). For ZSAC, we compare with all the baselines mentioned earlier as well as Wav2CLIP~\cite{wav2clip}, AudioClip~\cite{audioclip}, CoLLAT~\cite{silva2023collat} and ReCLAP w/ only desc. where we only employ $t$ acoustic properties as prompts and don't generate captions.
\vspace{1mm}

{\noindent \textbf{Ablations.}} We perform several ablations to prove the effectiveness of our approach. \textbf{For multi-modal retrieval:} (i) ReCLAP-660k: ReCLAP trained with caption augmentations on 660k pairs from ~\cite{ghosh2024compa}; (ii) CLAP-2.3M.: CLAP trained on our 2.3M audio-caption dataset (without caption augmentations). \textbf{For ZSAC:} For ZSAC we add another ablation which is: ReCLAP w/ only desc: ZSAC with ReCLAP with only the $t$ descriptions as prompts from the prompt augmentation stage and do not generate the $N$ diverse scenario prompts. 

\section{Results}
\label{sec:results}

Table~\ref{tab:exp-t2a-retrieval} compares ReCLAP to prior works on AudioCaps and Clotho for text-to-audio and audio-to-text retrieval, showing SOTA performance in most cases. ReCLAP-660k, trained on the same data as CompA-CLAP, surpasses it on all metrics, and similar results are seen for CLAP-2.3M vs. ReCLAP. This demonstrates that ReCLAP's improvements are not due to dataset size alone. Inspired by ~\cite{ghosh2024compa}, we argue that current benchmarks do not fully capture ReCLAP's capabilities in free-form T-A and A-T retrieval.


Table~\ref{tab:zshot_res} compares our prompt augmentation method (``w/ Prompt Aug'') with standard template-based prompting (``w/o Prompt Aug''). Prompt augmentation consistently outperforms baselines, with gains of 0.6\%-54.8\%. ReCLAP shows a 0.9\%-17.5\% improvement with prompt augmentation. In contrast, CLAP and LAION-CLAP show limited gains, indicating they don't interpret sound descriptions as effectively as ReCLAP. This highlights the importance of caption augmentation training as a prior step.



%


\section{Result Analysis}
\label{sec:result_analysis}

\begin{figure}[h]
    \centering
    \includegraphics[width=5.5cm,height=3.3cm]{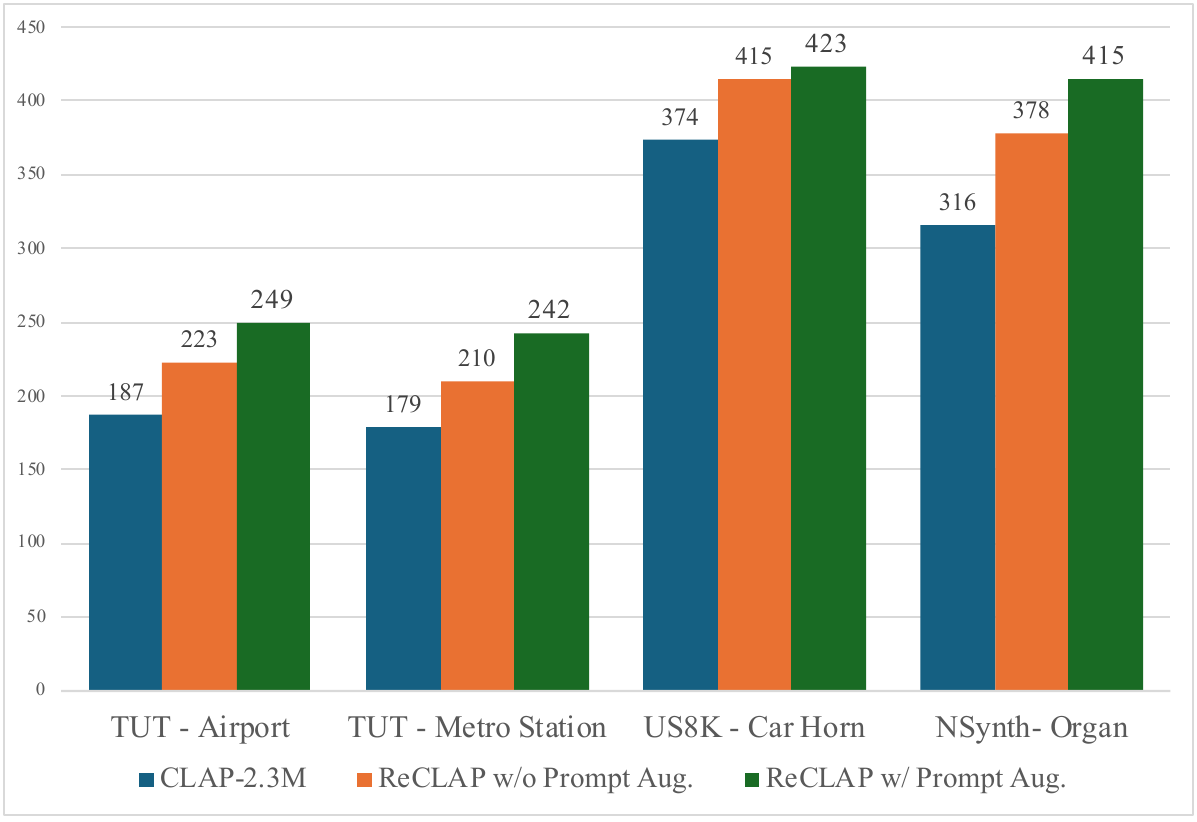}
    \caption{\small Comparison of accurately classified instances for 4 labels.}
    \label{fig:enter-label2}
\end{figure}

Fig.~\ref{fig:enter-label2} illustrates how ReCLAP and prompt augmentation increase the number of correct predictions for 4 labels from different datasets on ZSAC. As we can see, training CLAP on caption augmentation on the same dataset (CLAP-2.3M vs ReCLAP) improves retrieval of the correct label, which is further boosted by prompt augmentation. 

\begin{table}[h]
\caption{\small Examples of ambiguous labels where prompt augmentations provide additional context.}
\label{tab:examples}
\resizebox{1\columnwidth}{!}{
\begin{tabular}{ll}
\toprule \toprule
\textbf{Label}      & \textbf{Prompt Augmentation}  \\
\midrule
metro station (TUT) & A metallic rhapsody performed by the tireless locomotives, \\ & with a recurring refrain from the station's vocal spirit. \\ \midrule
airport (TUT) & A chorus of engines hums persistently, interspersed with the\\ & murmur of voices and authoritative announcements. \\ \midrule
organ (NSynth) & The organ unfurled a tapestry of majestic harmonies, filling \\ & the cathedral with its thunderous hymn.           \\ \midrule
writing (FSD50K) & In the hush of the library, the rhythmic scratch of pen on\\ &paper becomes a soft dance of intellect and ink.   \\ \bottomrule
\end{tabular}}
\end{table}
For example, the sound of an "organ" could refer to either a human organ or a musical instrument. By adding useful contextual information, prompt augmentations assist in clarifying such ambiguities, leading to more accurate retrievals.

\section{Hyper-parameter Tuning}
Table~\ref{tab:number_n} compares performance across $N$=\{1,2,3,4,5\} to show the effect of the number of custom prompts $N$ on the final ZSAC performance. As we see, the optimal performance is achieved at $N$=2, and model performance decreases with an increase in $N$. This decline is hypothesized to be due to the introduction of more noise into the process with each additional caption.

\begin{table}[h]
\caption{\small Impact of $N$ on ZSAC with ReCLAP.}
    \centering
    \resizebox{0.75\columnwidth}{!}{
    \begin{tabular}{@{}c|lllll@{}} 
    \hline \hline
$N$     & 1 & 2 & 3 & 4 & 5 \\ \hline
Score   & 48.56  & \textbf{52.22}  & \underline{49.37}  & 47.24  & 44.35 \\ \hline
    \end{tabular}}
    
    \label{tab:number_n}
\end{table}

Additionally, Table~\ref{tab:prob} shows the effect of probability $p$ on the final ZSAC performance. 

\begin{table}[h]
\caption{\small Impact of $p$ on ZSAC with ReCLAP.}
    \centering
    \resizebox{0.75\columnwidth}{!}{
    \begin{tabular}{@{}c|lllll@{}}
    \hline \hline
$p$     & 0.2 & 0.4 & 0.6 & 0.8 \\ \hline
Score & 47.16    & \textbf{52.22}    & \underline{48.29}    & 45.64 \\ \hline
    \end{tabular}}
    \label{tab:prob}
\end{table}

\section{Conclusion}
\label{sec:conclusion}

In this paper, we propose to improve ZSAC by interpreting sounds using their descriptive features. To achieve this, we first propose ReCLAP, a CLAP model trained using additional caption augmentations that improve CLAP's understanding of sounds in the wild. Next, we propose to improve ZSAC with prompt augmentation, where we move beyond standard hand-written prompts and generate custom prompts for each category in the dataset. Our proposed method improves ZSAC over our baselines by significant margins.

\section{Limitations and Future Work}
\label{sec:limitations}

\begin{enumerate}

    \item LLM-generated augmentations may result in errors or repetitive captions, which require substantial human oversight. Future work will explore methods to improve quality control.

    \item Synthetic augmentations from LLMs may introduce biases into models. Future efforts will focus on mitigating these biases.

    \item Representations from ReCLAP can be employed to improve a range of tasks, including audio generation and understanding. Future work includes exploring these tasks.
\end{enumerate}

\bibliography{reclap}
\bibliographystyle{bib}

\end{document}